\documentclass[aps,pra,preprintnumbers,superscriptaddress,nofootinbib,12pt,notitlepage]{revtex4-1}
\usepackage{siunitx}
\usepackage{graphicx}
\usepackage{setspace}
\usepackage{diagbox}
\usepackage{xcolor}
\usepackage{slashed}
\usepackage{amsmath,amssymb}
\usepackage{makecell}
\usepackage{floatrow}

\begin{document}

\preprint{FERMILAB-PUB-20-378-QIS-T}

\title{Mechanical Quantum Sensing in the Search for Dark Matter}
\author{{\bf Conveners:} D. Carney}
\thanks{carney@umd.edu}
\affiliation{Joint Center for Quantum Information and Computer Science/Joint Quantum Institute, 
University of Maryland/NIST, College Park/Gaithersburg, MD, USA}
\affiliation{Fermi National Accelerator Laboratory, Batavia, IL 60510, USA}
\author{G. Krnjaic}
\thanks{krnjaicg@fnal.gov}
\affiliation{Fermi National Accelerator Laboratory, Batavia, IL 60510, USA}
\affiliation{Kavli Institute for Cosmological Physics and Enrico Fermi Institute, University of Chicago, Chicago, IL 60637, USA}
\author{D. C. Moore}
\thanks{david.c.moore@yale.edu}
\affiliation{Wright Laboratory, Department of Physics, Yale University, New Haven, CT, USA}
\author{C. A. Regal}
\thanks{regal@colorado.edu}
\affiliation{JILA, National Institute of Standards and Technology/University of Colorado, Boulder, CO 80309, USA}
\affiliation{Department of Physics, University of Colorado, Boulder, CO 80309, USA}

\author{ \\ \vspace{8pt} G. Afek}
\affiliation{Wright Laboratory, Department of Physics, Yale University, New Haven, CT, USA}
\author{S. Bhave}
\affiliation{Department of Electrical and Computer Engineering, Purdue University, West Lafayette, IN 47907, USA}
\author{B. Brubaker}
\affiliation{JILA, National Institute of Standards and Technology/University of Colorado, Boulder, CO 80309, USA}
\affiliation{Department of Physics, University of Colorado, Boulder, CO 80309, USA}
\author{T. Corbitt}
\affiliation{Department of Physics and Astronomy, Louisiana State University, Baton Rouge, LA 70803, USA}
\author{J. Cripe}
\affiliation{National Institute of Standards and Technology, Gaithersburg, MD 20899, USA}
\author{N. Crisosto}
\affiliation{Department of Physics, University of Washington, Seattle, WA 98195, USA}
\author{A. Geraci}
\affiliation{Department of Physics and Astronomy, Northwestern University, Evanston, Illinois 60208, USA}
\author{S. Ghosh}
\affiliation{Joint Center for Quantum Information and Computer Science/Joint Quantum Institute, 
University of Maryland/NIST, College Park/Gaithersburg, MD, USA}
\author{J. G. E. Harris}
\affiliation{Wright Laboratory, Department of Physics, Yale University, New Haven, CT, USA}
\author{A. Hook}
\affiliation{Maryland Center for Fundamental Physics, Department of Physics,
University of Maryland, College Park, MD 20742, USA}
\author{E. W. Kolb}
\affiliation{Kavli Institute for Cosmological Physics and Enrico Fermi Institute, University of Chicago, Chicago, IL 60637, USA}
\author{J. Kunjummen}
\affiliation{Joint Center for Quantum Information and Computer Science/Joint Quantum Institute, 
University of Maryland/NIST, College Park/Gaithersburg, MD, USA}
\author{R. F. Lang}
\affiliation{Department of Physics and Astronomy, Purdue University, West Lafayette, IN 47907, USA}
\author{T. Li}
\affiliation{Department of Electrical and Computer Engineering, Purdue University, West Lafayette, IN 47907, USA}
\affiliation{Department of Physics and Astronomy, Purdue University, West Lafayette, IN 47907, USA}
\author{T. Lin}
\affiliation{Department of Physics, University of California, San Diego, CA 92093, USA}
\author{Z. Liu}
\affiliation{Maryland Center for Fundamental Physics, Department of Physics,
University of Maryland, College Park, MD 20742, USA}
\author{J. Lykken}
\affiliation{Fermi National Accelerator Laboratory, Batavia, IL 60510, USA}
\author{L. Magrini}
\affiliation{Vienna Center for Quantum Science and Technology, Faculty of Physics, University of Vienna, A-1090 Vienna, Austria}
\author{J. Manley}
\affiliation{Department of Electrical and Computer Engineering, University of Delaware, Newark, DE 19716, USA}
\author{N. Matsumoto}
\affiliation{Research Institute of Electrical Communication, Tohoku University, Sendai 980-8577, Japan}
\affiliation{Frontier Research Institute for Interdisciplinary Sciences, Tohoku University, Sendai 980-8578, Japan}
\affiliation{JST, PRESTO, Kawaguchi, Saitama 332-0012, Japan}
\author{A. Monte}
\affiliation{Fermi National Accelerator Laboratory, Batavia, IL 60510, USA}
\author{F. Monteiro}
\affiliation{Wright Laboratory, Department of Physics, Yale University, New Haven, CT, USA}
\author{T. Purdy}
\affiliation{Pittsburgh Quantum Institute, University of Pittsburgh, Pittsburgh, PA 15260, USA}
\author{C. J. Riedel}
\affiliation{NTT Research Inc., Physics \& Informatics Laboratories, Sunnyvale, CA, USA}
\author{R. Singh}
\affiliation{National Institute of Standards and Technology, Gaithersburg, MD 20899, USA}
\author{S. Singh}
\affiliation{Department of Electrical and Computer Engineering, University of Delaware, Newark, DE 19716, USA}
\author{K. Sinha}
\affiliation{Department of Electrical Engineering, Princeton University, Princeton, NJ 08544, USA}
\author{J. M. Taylor}
\affiliation{Joint Center for Quantum Information and Computer Science/Joint Quantum Institute, 
University of Maryland/NIST, College Park/Gaithersburg, MD, USA}
\author{J. Qin}
\affiliation{Department of Physics and Astronomy, Purdue University, West Lafayette, IN 47907, USA}
\author{D. J. Wilson}
\affiliation{Wyant College of Optical Sciences, University of Arizona, Tucson, AZ 85721, USA}
\author{Y. Zhao}
\affiliation{Department of Physics and Astronomy, University of Utah, Salt Lake City, UT 84112, USA}
\date{\today}

\singlespace

\begin{abstract}
 
Numerous astrophysical and cosmological observations are best explained by the existence of dark matter, a mass density which interacts only very weakly with visible, baryonic matter. Searching for the extremely weak signals produced by this dark matter strongly motivate the development of new, ultra-sensitive detector technologies. Paradigmatic advances in the control and readout of massive mechanical systems, in both the classical and quantum regimes, have enabled unprecedented levels of sensitivity.  In this white paper, we outline recent ideas in the potential use of a range of solid-state mechanical sensing technologies to aid in the search for dark matter in a number of energy scales and with a variety of coupling mechanisms. 
 
\end{abstract}

\maketitle

\tableofcontents 

\newpage

\section{Introduction}

A significant and growing body of astrophysical \cite{Sofue:2000jx,Markevitch:2003at,Massey:2010hh} and cosmological \cite{Primack:2015kpa,Aghanim:2018eyx} observations strongly suggests the existence of ``dark matter'', a massive substance which interacts very weakly---perhaps only through gravity---with ordinary, visible matter. This dark matter has not yet been observed at particle colliders or in dedicated searches \cite{RevModPhys.90.045002}. Many dark matter direct detection experiments to date have focused on weakly interacting massive particles (WIMPs) with masses around $100~\rm{GeV}$. These technologies are reaching full maturity, and will have either detected or largely excluded WIMPs as viable dark matter candidates within the next generation of experiments \cite{arcadi2018waning}. There is thus a clear need for searches of new dark matter candidates, with new experimental techniques \cite{battaglieri2017us}. 

Precision measurement techniques have already been deployed in the search for dark matter (see e.g.\ \cite{Ahmed:2018oog,RevModPhys.90.025008} for reviews).  In this white paper, we discuss approaches to searching for dark matter using massive, mechanical sensing devices.  We include applications of purely classical mechanical sensors, as well as many devices which are now operating in the ``quantum-limited'' regime, in which the dominant noise contributions come from the quantum mechanics of measurement itself. These ultra-high precision systems can enable tests of a wide range of dark matter models with extremely small couplings to ordinary matter (both electromagnetic and otherwise). These approaches complement existing search strategies, and in many cases provide better sensitivity than other available options. 

The development of mechanical detectors has a rich history.  Precision measurement in the context of gravitational physics has utilized a range of large-scale systems such as optical interferometers~\cite{abramovici1992ligo}, atom interferometers~\cite{kasevich1992measurement,peters2001high}, torsion balances~\cite{hoyle2001submillimeter,wagner2012torsion}, and Weber bars~\cite{weber1966observation,cerdonio1997ultracryogenic}. The broader landscape of study of mechanical systems, as both classical and quantum detectors, is wide reaching--ranging from single ions \cite{biercuk2010ultrasensitive,ivanov2016high}, to tens of thousands of atoms \cite{schreppler2014optically}, to microscale resonators \cite{teufel2009nanomechanical,peterson2016laser} and up to kilogram-scale devices \cite{abramovici1992ligo,hoyle2001submillimeter}.  In this white paper, we consider how a variety of mechanical systems can open fundamentally new avenues to search for dark matter over a large range of energy scales. In particular, monitoring solid, massive objects allows for coherent integration of long-wavelength interactions, and for integration of small cross sections over large volumes or large numbers of target atoms or nuclei. Mechanical devices that are read out interferometrically at the shot-noise limit, or even at or below the standard quantum limit (SQL) enforced by quantum backaction \cite{caves1980quantum}, have been demonstrated across a wide range of mass scales, with natural frequencies ranging from millihertz to terahertz in recent years (see \cite{aspelmeyer2014cavity} for a review). Hence, multiple technologies are at an opportune point for contemplating their role in precision experiments.

Dark matter detection is a particularly compelling and challenging problem, which may require the development of fundamentally new technologies. Mechanical detection may be poised to contribute to these challenging searches in both near-term and long-term experiments. Development of new technologies will necessarily proceed with researchers in the sensing and particle physics communities working in tandem. In the following, we outline opportunities and objectives in this new direction in the search for dark matter. We note that the mechanical sensing techniques we focus on have many similarities to proposed dark matter searches with atom interferometry \cite{graham2016dark,geraci2016sensitivity,Coleman:2018ozp} and atomic clock systems \cite{derevianko2014hunting,arvanitaki2015searching,stadnik2015searching}, but here we focus on the domain of solid objects.  

\section{Motivations for mechanical sensors}
\label{section-mass}

The present landscape of viable dark matter candidates is enormous, leading to a wide variety of potential experimental signatures. Dark matter particles could range in mass from $10^{-22}~{\rm eV}$ up to hundreds of solar masses, a range of some 90 orders of magnitude.\footnote{In this paper, we use natural units $\hbar = c = 1$ to quote particle physics quantities like masses and momenta.} Moreover, dark matter could interact with the standard model through many possible interactions, although perhaps only through gravity. To span this diverse range of possible models, different regions of parameter space will require different detector architectures and measurement techniques. In particular, for models interacting with the standard model only through mass or other extensive quantities such as nucleon number, massive mechanical sensors may be required. Mechanical sensing technologies offer an extensive set of platforms, as discussed in section~\ref{section-detectors}, and thus have the potential to search for a wide range of such dark matter candidates in regions of parameter space that are complementary to existing searches.

The ability to monitor a large number of atoms in aggregate offers two key advantages over other approaches. The first advantage is the large volume integration of any putative dark matter signal. Any dark-visible interactions are necessarily tiny, so using a large volume (or a large mass of target nuclei or atoms, for models that can resolve the underlying substructure of the masses) is key to meaningful detection prospects. The second advantage is that long-wavelength signals can be integrated coherently across the full device, leading to greatly enhanced sensitivities. Such coherent detection has applications in the search for signals from wave-like dark matter signals like the axion or other ultralight bosons, as well as in the case of impulses delivered with extremely small momentum transfers. In section~\ref{section-applications}, we give some examples of dark matter models leading to these types of signals, and discuss prospects for their detection with mechanical sensors.

\section{Detection targets and techniques}
\label{section-applications}

\begin{figure}[t!]
    \centering
    \includegraphics[scale=0.4]{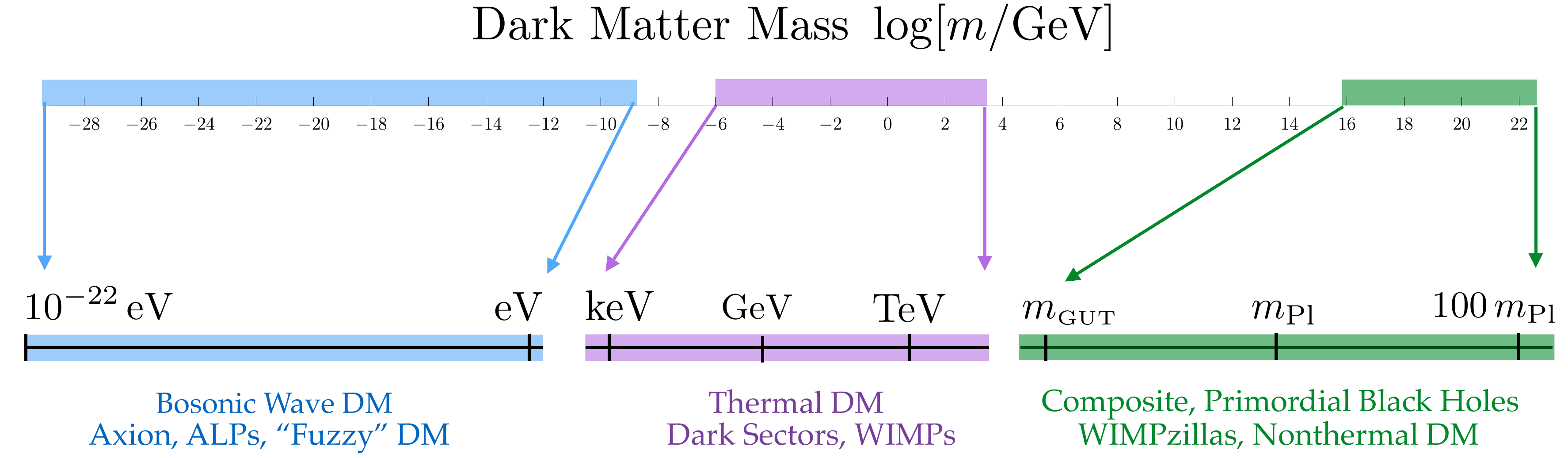}
    \caption{\emph{Range of available dark matter candidates}. Current observations allow for dark matter to consist of quanta with an enormous range of masses. Here we classify these candidates as particle-like when $m \gtrsim 1~{\rm eV}$, and ultralight, wave-like dark matter when $m \lesssim 1~{\rm eV}$. A few prototypical models are listed as examples.}
    \label{figure-massrange}
\end{figure}

Possible signals of dark matter are controlled by a few key parameters. Astrophysical observations tell us that the dark matter mass density in our neighborhood is $\rho \sim ~0.3~{\rm GeV/cm^3}$ \cite{read2014local}. Assuming this dark matter consists of a single component, with (unknown) mass of an individual dark matter quantum, $m_{\chi}$, this means that the local number density is around
\begin{equation}
\label{nchi}
n_{\chi} = \frac{0.3}{\rm cm^3} \times \left( \frac{1~{\rm GeV}}{m_{\chi}} \right).
\end{equation}
Moreover, the Earth is moving through the virialized background dark matter with ``wind speed'' $v_{DM} \sim 200~{\rm km/s}$. These parameters fix the kinematics of any detection experiment. The only additional information is what non-gravitational couplings, if any, the dark matter has with visible matter. See eg. \cite{Lin:2019uvt} for a review and further references.

Broadly speaking, the above properties mean that potential dark matter signals fall into two classes determined by the dark matter particle mass (see Fig.~\ref{figure-massrange}). Traditional DM detection has focused on dark matter candidates of masses greater than around $m_{\chi} \gtrsim 1~{\rm eV}$, which appear as distinct particles. If these interact with visible matter, they will deposit tiny, discrete impulses (on the order of $p = m_{\chi} v_{DM}$) when they collide with a detector. On the other hand, ultralight dark matter fields of mass $10^{-22}~{\rm eV} \lesssim m_{\chi} \lesssim 1~{\rm eV}$ have enormous occupation numbers, given Eqn.~\eqref{nchi}. The low mass and high occupation number of the quanta mean that the field is bosonic and behaves as a background of oscillating waves of wavelength $\lambda_{\rm dB} \gtrsim 1~{\rm mm}$. This background of waves will be coherent over a timescale $T_{\rm coh} \sim 10^6/\omega_{\chi}$ set by Doppler broadening, where $\omega_{\chi} = m_{\chi} c^2/\hbar$ is the natural frequency of the field~\cite{sikivie1983experimental,hu2000fuzzy}. These models thus produce extremely weak, coherent, persistent signals. Searching for these two classes of signals requires different measurement techniques, which we now discuss separately in more detail.

\subsection{Ultralight searches}

\begin{figure}[t!]
    \centering
    \includegraphics[scale=0.38]{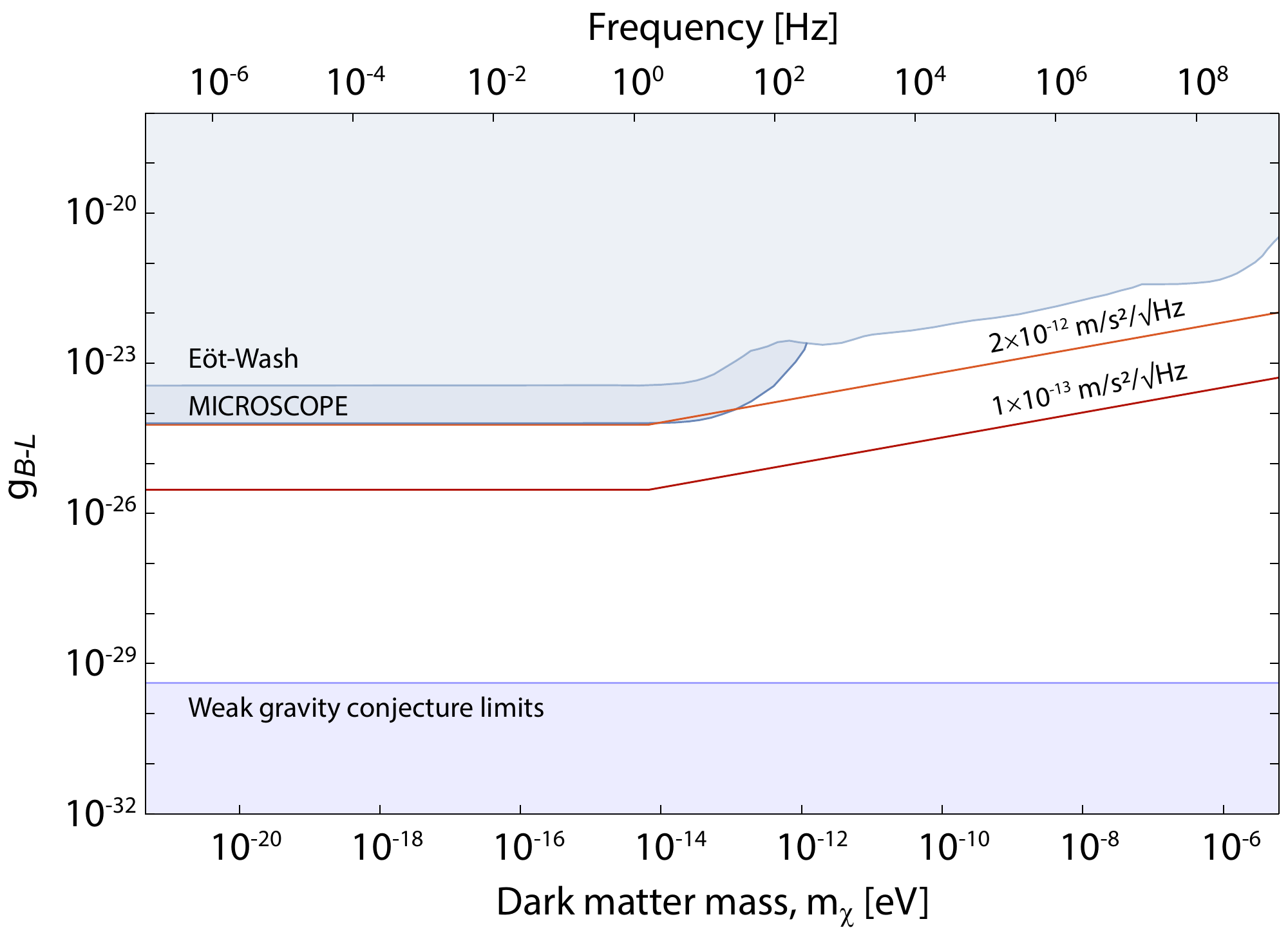} ~ \includegraphics[scale=0.38]{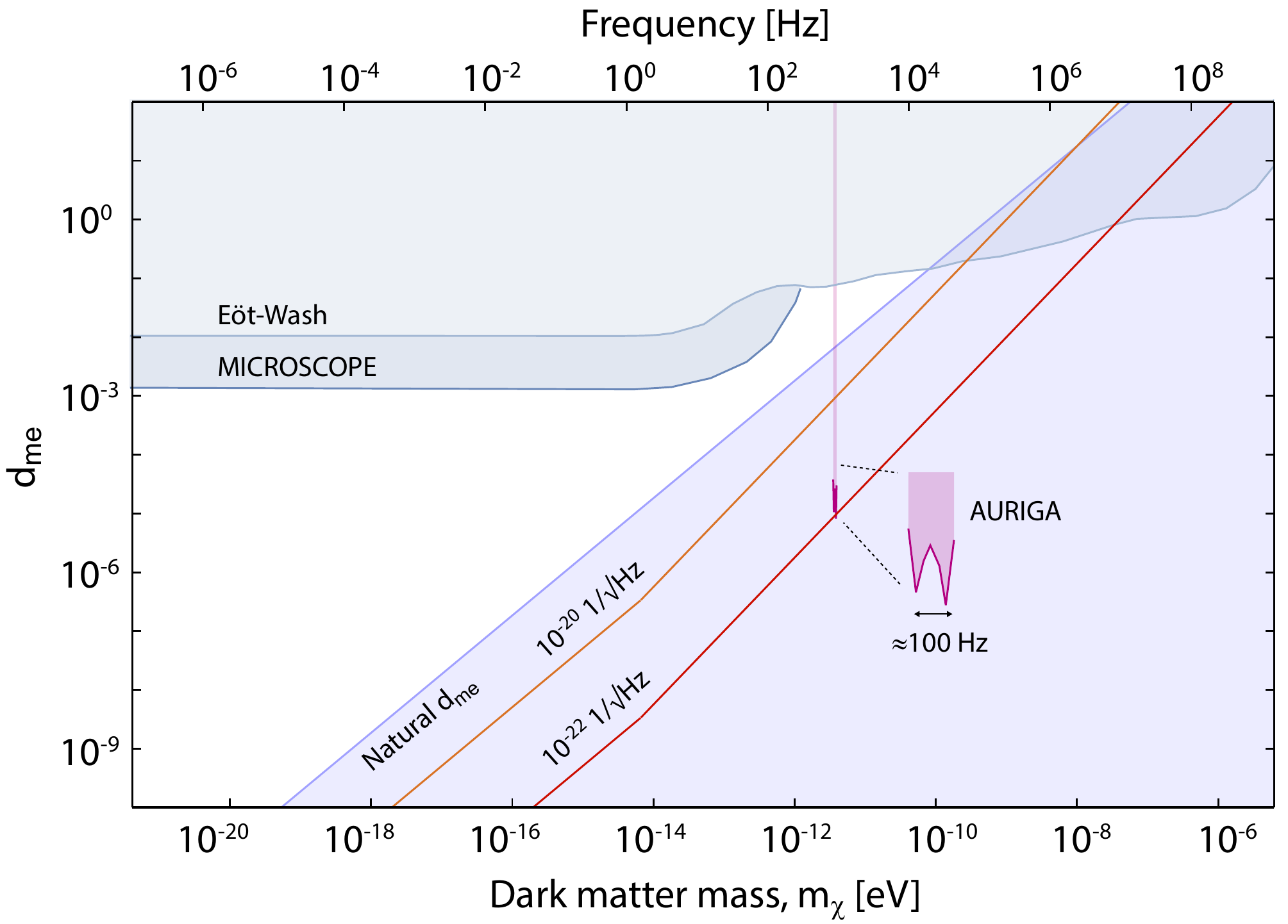}
    \caption{\emph{Ultralight dark matter searches}. Left: Detection reach for accelerometer searches of ultralight dark matter \cite{graham2016dark,Carney:2019cio}, taking a vector $B-L$ boson as an example. We assume one day of integration time, and the use of a pair of accelerometers with differential neutron-to-nucleon ratio $\Delta = N_1/A_1 - N_2/A_2 = 0.05$. Upper shaded regions are ruled out by existing torsion-balance \cite{schlamminger2008test,wagner2012torsion,arvanitaki2016sound} and satellite experiments \cite{hees2018violation,berge2018microscope}. Right: Detection reach for strain sensors \cite{arvanitaki2016sound,manley2019searching}, using a scalar field coupled to electrons as an example. The AURIGA Weber bar experiment provides an additional narrow-band constraint \cite{branca2017search}. In both plots, the colored lines labeled by sensitivities represent the lower limit of dark matter parameter space which can be probed with a detector of the given sensitivity. The lower shaded regions give some examples of conjectural theory input: the region in the left plot conflict with a version of the weak gravity conjecture  \cite{ArkaniHamed:2006dz,cheung2018proof}, here applied assuming the lightest $B-L$ coupled particle is a neutrino of mass $0.01~{\rm eV}$. In the right plot, the lower shaded region is favored by naturalness arguments \cite{arvanitaki2016sound}.}
    \label{figure-ultralightplot}
\end{figure}
Consider a scenario where a sizeable fraction of the dark matter mass density is made up of a single ultralight field. Examples of such ultralight dark matter candidates include the axion~\cite{sikivie1983experimental}, vector bosons arising by gauging the conservation of baryon minus lepton number ($B-L$) \cite{graham2016dark}, scalar and pseudoscalar fields coupled through the Higgs portal \cite{piazza2010sub} or the stress tensor \cite{arvanitaki2015searching} (see Table 1 of Ref.~\cite{graham2016dark} for a collation of allowed couplings). These models are minimal in the sense that they add only a single field to the standard model of particle physics, and introduce no ultraviolet anomalies. The axion couples directly to the electromagnetic and gluon fields, and can thus be searched for using a variety of systems including microwave cavities \cite{du2018search,zhong2018results} and NMR systems \cite{arvanitaki2014resonantly}. The other candidates, however, can couple to quantities proportional to mass density. It is thus natural to search for these types of DM with massive sensors.

If DM consists primarily of one of these ultralight fields, the observable signature is an oscillating background of ultralight bosons. This produces a nearly monochromatic, sinusoidal force signal in a massive detector, with strength proportional to the mass, leading to a variety of physical effects. For scalar DM the variations of fundamental constants such as the electron mass, or fine structure constant would lead to a periodic strain in macroscopic devices, and the possibility of detecting it has been explored in several mechanical structures \cite{arvanitaki2016sound,branca2017search,manley2019searching,geraci2019searching}. For pseudoscalar DM candidates, observable signatures can include time-varying nucleon electric dipole moments, spin-torques, and EMFs along magnetic fields \cite{graham2016dark}. For vector DM one can obtain material dependent couplings, leading to differential accelerations. For a concrete example, consider a vector boson field $A_{\mu}$ arising from a gauged $B-L$ symmetry. This couples to the neutron field $n$ through the neutron number density, that is, through a coupling $g_{B-L} \slashed{A} \overline{n} n$. The dark matter background of vector bosons then leads to a force on a sensor given by
\begin{equation}
\label{uldmsignal}
    F(t) = F_0 N_n g_{B-L} \cos(m_{\chi}c^2 t/\hbar)
\end{equation}
where $N_n$ is the number of neutrons in the sensor, $F_0 \sim 10^{-15}~{\rm N}$ is set by the dark matter density \eqref{nchi}, and $g_{B-L}$ is an unknown but weak coupling strength \cite{graham2016dark,Carney:2019cio}. Since the coupling is to neutron number as opposed to total mass, a pair of sensors with different neutron-to-nucleon ratios $N/A$ can be used to search for the differential acceleration produced by \eqref{uldmsignal}. In Fig. \ref{figure-ultralightplot}, we plot the available parameter space in this scenario and the acceleration sensitivities needed for novel searches.

At the core, the detection problem here is to sense a weak, persistent, narrow-band signal. Coherent sensing of narrowband forces is a prototypical application of mechanical sensors, and so these are ideal detection targets for which mechanical sensors are poised to make an immediate impact, particularly at higher frequencies (Hz-GHz) and/or using multiple sensors to coherently integrate the signal. 

\subsection{Particle-like searches based on recoils}

To detect heavier ($m_{\chi} \gtrsim 1~{\rm eV}$), particle-like dark matter candidates, a variety of techniques can be used. The key challenges in this regime can be illustrated by reviewing traditional WIMP detection (see Ref.~\cite{Schumann:2019eaa} for a review). In a liquid noble detector, the WIMPs would occasionally strike an atomic nucleus, causing it to recoil. If sufficient energy was deposited, the nucleus ionizes or excites nearby atoms, leading to either electron-ion pairs or emission of scintillation photons which can then be detected by charge sensors or photodetectors at the edges of the detector. This example demonstrates the basic issues: the events are very rare (owing to the tiny dark matter-nucleon cross sections, $\sigma \lesssim 10^{-36}~{\rm cm^2}$ \cite{Akerib:2016vxi}) and the energy deposition is very small (a given WIMP has mass of about $\sim$100 protons and velocity $10^5~{\rm m/s}$) leading to only small amounts of ionization or scintillation. Thus any detection program needs to have sufficient target mass to see enough events, as well as very low detection thresholds to see these small energy deposits. We note that many other signals of interest, in particular low-energy neutrinos \cite{cabrera1985bolometric}, have precisely the same properties.

The massive mechanical sensing paradigm offers a straightforward solution to the issue of mass: for example, the LIGO detectors have mechanical elements (the interferometer mirrors) with masses of tens of kilograms! On the other hand, smaller mechanical detectors can also enable extremely low-threshold energy detection. There are two basic strategies: detection of localized phonons in bulk materials, and direct monitoring of impulses to the center of mass motion of a single device.

\begin{figure}[t]
\floatbox[{\capbeside\thisfloatsetup{capbesideposition={left,top},capbesidewidth=.5\linewidth}}]{figure}[\FBwidth]
{\caption{Schematic of a phonon-counting experiment with liquid helium in an optomechanical cavity \cite{shkarin2019quantum}. Darker blue indicates superfluid helium, light blue is glass. Blue shading indicates a typical paraxial acoustic mode, and the red shows the optical mode to which it couples. Optical modes with wavelength $1550~{\rm nm}$ couple to acoustic modes with frequency $315~{\rm MHz}$, corresponding to energies around $1.5~{\rm \mu eV}$. An excited phonon mode can convert into an off-resonance photon through a Stokes or anti-Stokes process. By filtering out the resonant photons, this enables counting of the phonon excitations with temporal resolution set by the photodetector (here on the order of $50~{\rm ns}$). In this example, the fluid is held at a temperature $25~{\rm mK}$ and individual thermal phonons are being counted. These phonons can be cooled out of the cavity mode, to enable detection of athermal phonons (as e.g.\ produced by dark matter collisions with the helium).}  \label{figure-phonons}}
{\includegraphics[width=.98\linewidth]{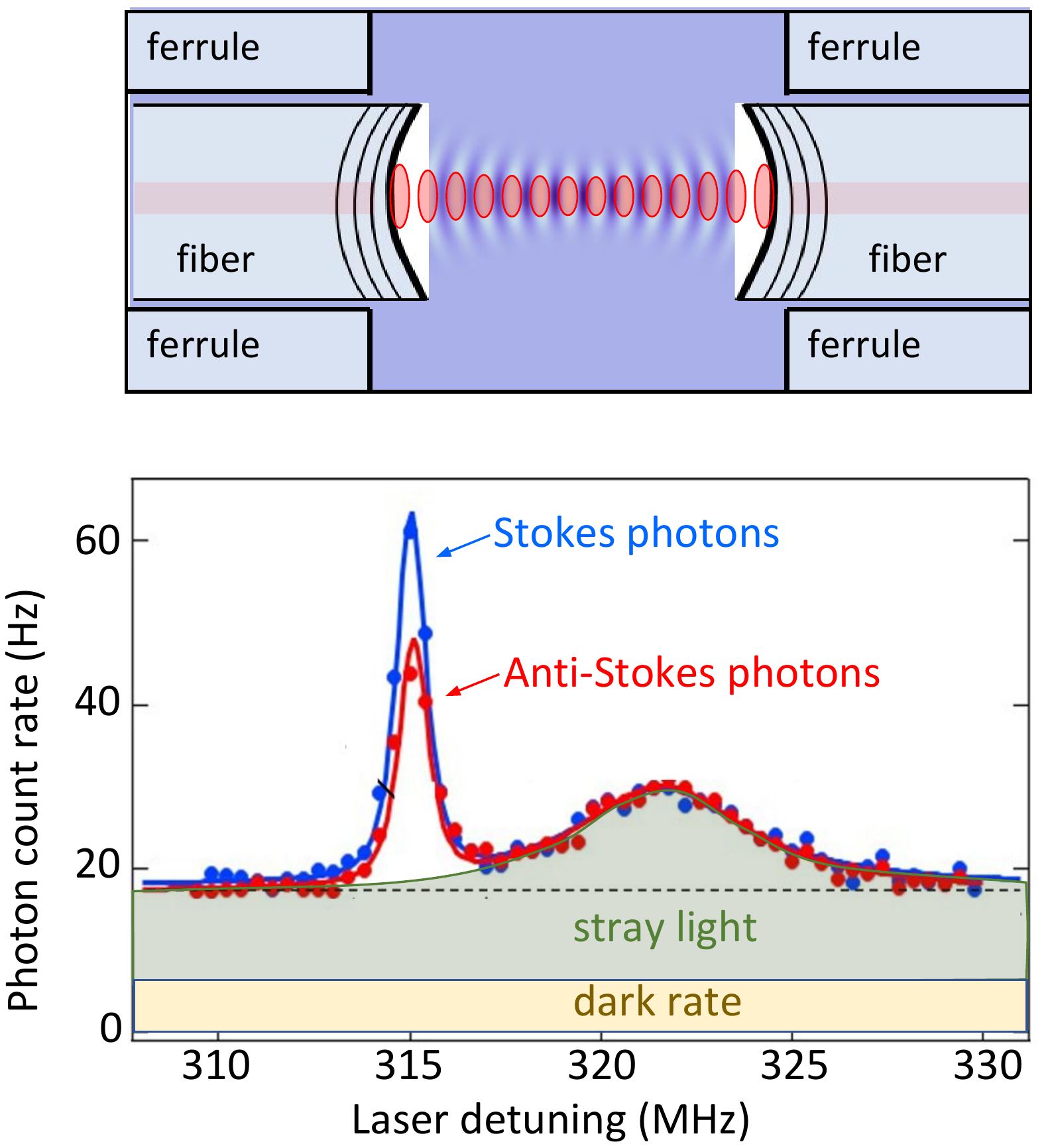}}
\end{figure}

A number of proposals for the detection of dark matter through bulk phononic excitations currently exist \cite{guo2013concept,Schutz:2016tid,griffin2018directional,knapen2018detection,Kurinsky:2019pgb}, which may extend the sensitivity beyond existing implementations of phonon sensing in cryogenic calorimeters (e.g.~\cite{SuperCDMSSNOLAB:2017,CRESST:2019,Edelweiss:2019}). For example, when a dark matter particle interacts with a nucleus in a bulk crystal, it generates a distortion of the lattice. In particular, if the inverse momentum transfer is larger than the lattice spacing, phonons are excited. The phonons then travel through the material, and can be sensed by calorimetric detectors at the edges of the material. As an example, state-of-the-art transition edge sensors can resolve a total deposited energy in phonons down to energies around ${\rm few} \times 10 ~{\rm meV}$ \cite{Fink:2020noh}. This means that searches of this type are sensitive to ``light'' dark matter candidates, of masses in the eV-MeV range. Optomechanical readout of phonons in small samples can reach substantially lower thresholds. For example, single phonons at the micro-eV level can be read out in micromechanical oscillators \cite{cohen2015phonon,riedinger2016non} superfluid helium~\cite{shkarin2019quantum} or bulk crystals \cite{jain2020listening}; we show the superfluid helium example in Fig.~\ref{figure-phonons}. The primary challenge in such systems is not energy threshold, but instead coupling energy into the phonon modes of interest (which are often purposefully decoupled from the bulk phonon modes in the system to avoid thermal noise).  In addition, such systems are small (with mode masses at the $\mu$g to mg scale), so scaling up to a sufficient volume for non-trivial dark matter detection reach is an interesting open problem. If coupling of phonons into the modes of interest could be engineered (even with relatively low efficiencies) such techniques would provide an exciting complement to calorimetric phonon detection experiments.

\begin{figure}[t!]
    \centering
    \includegraphics[height=4.8cm]{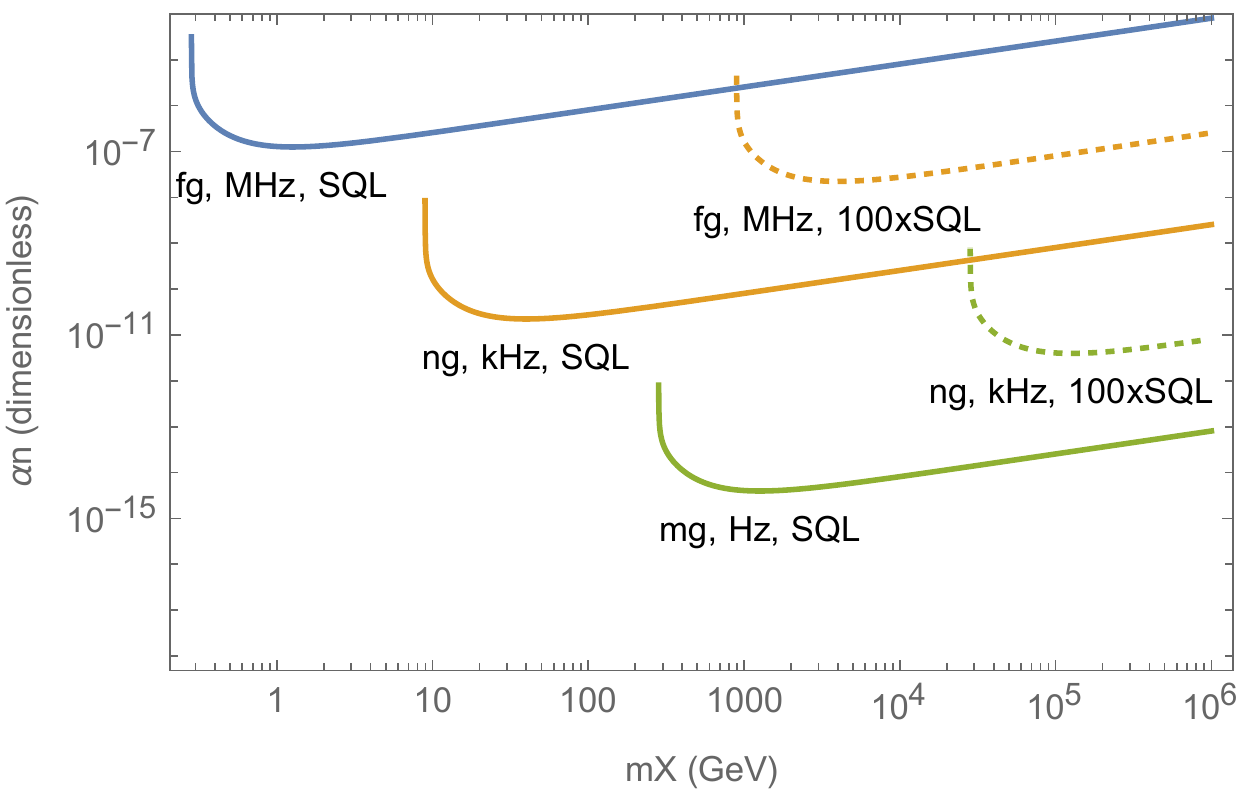}
    \includegraphics[height=4.8cm]{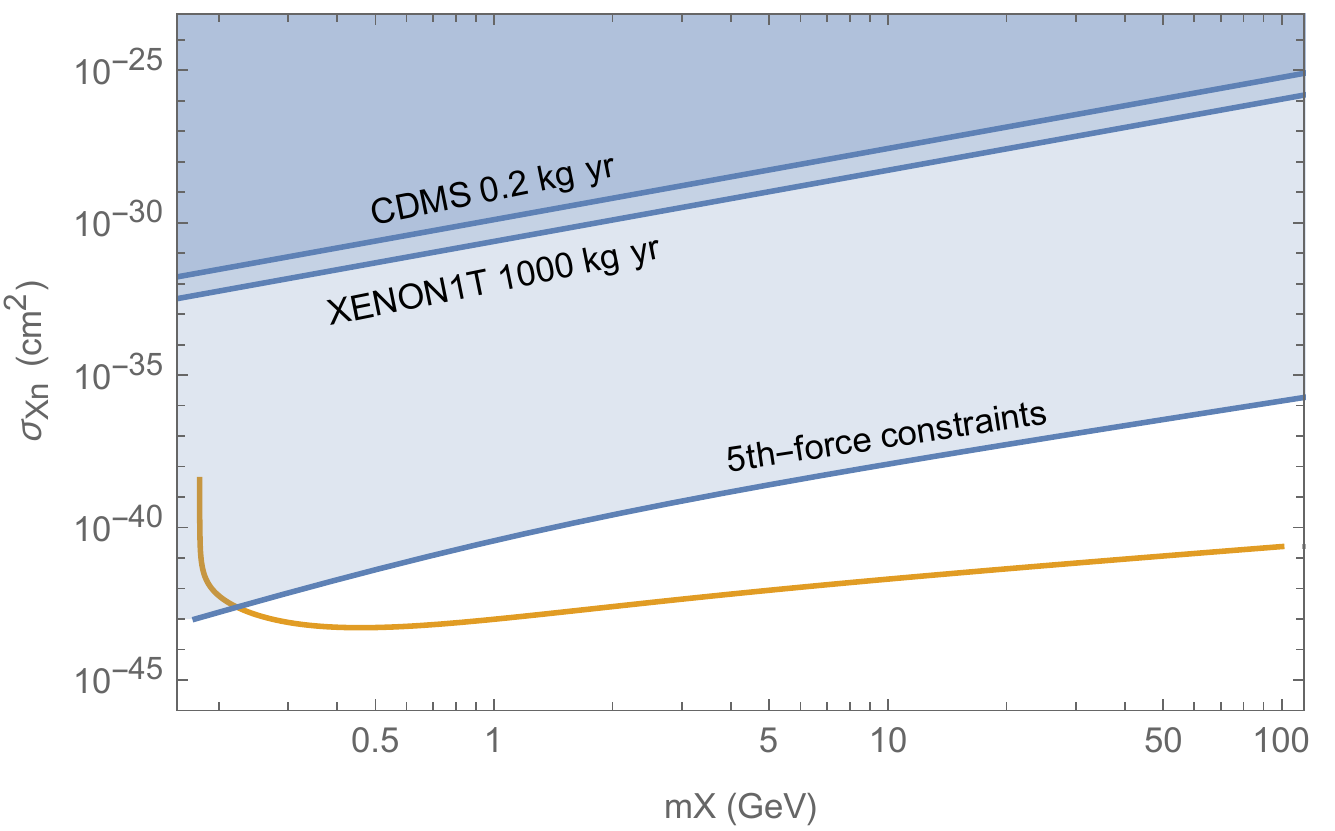}
    \caption{\emph{Searches for particle-like dark matter}. Here we consider dark matter consists primarily of particles of mass $m_X$, coupling to neutrons through a light mediator (eg. through a potential $V = \alpha_n/r$, where $\alpha_n$ is a small, unknown coupling strength) as an example search target for mechanical impulse sensors. In the left plot, each curve represents a hypothetical sensor (labeled by its mass, readout frequency, and noise level benchmarked to \eqref{SQL}). Sensitivity is lost at low mass because the incoming DM will not have enough momentum to deliver to the device, and at high mass because of the loss of flux (see Eqn.~\eqref{nchi}). In the right plot, we use a nanogram-scale sensor operated at the SQL as an example and show projected constraints compared to currently-existing bounds. To draw the current bounds, we assume a microscopic realization in which dark matter consists of ``nuggets'' of total mass $m_X$ made of multiple constituents of mass $m_{\chi} \sim 1~{\rm MeV}$, coupled to neutrons through a $B-L$ vector boson of mass $m_{\phi} \sim 0.05~{\rm eV}$ (for discussion of the parametrization of the fiducial DM-nucleon cross section $\sigma_{Xn}$, see Ref.~\cite{coskuner2019direct,Monteiro2020compositeDM}). The XENON1T \cite{Aprile:2017iyp} and CDMS \cite{Agnese:2017jvy} bounds come from pre-existing particle physics experiments while the fifth-force bounds come from torsion-balance searches \cite{schlamminger2008test,wagner2012torsion,arvanitaki2016sound,heeck2014unbroken}.}
    \label{figure-impulse}
\end{figure}

Alternatively, one can monitor the center of mass motion of an entire object (i.e. the zero-mode phonon). This technique could be particularly advantageous in the setting where the collision acts coherently on the entire mechanical component, for example when the dark matter couples to the sensor through a long-range force. Here one continuously monitors the center of mass position and looks for small transfers of momenta greater than the typical noise on the device. The noise floor is ultimately limited by thermal coupling with the environment and by quantum mechanical measurement noise coming from the monitoring of the device \cite{caves1980quantum,caves1981quantum}. Concretely, the standard quantum limit (SQL) provides a benchmark for a detectable impulse \cite{mozyrsky2004quantum,clerk2004quantum}: 
\begin{equation}
\label{SQL}
\Delta p_{\rm SQL} = \sqrt{\hbar m \omega} \approx 1.5~{\rm MeV} \times \left( \frac{m}{1~{\rm ng}} \right)^{1/2} \left( \frac{\omega/2\pi}{1~{\rm kHz}} \right)^{1/2},
\end{equation}
where $m, \omega$ are the mass and frequency of the mechanical sensor.\footnote{Here, the frequency $\omega$ should be replaced by the inverse measurement times scale when this exceeds the mechanical frequency, such as the free-mass case $\omega\to 0$.} While methods exist to go below this noise level (see Sec.~\ref{section-detectors}), currently existing devices acting at or even slightly above the SQL are already capable of searching novel regions of DM parameter space, as demonstrated by the initial search in~\cite{Monteiro2020compositeDM}. We describe an example in Fig. \ref{figure-impulse}.

\subsection{Direct gravitational interaction with particle-like dark matter}
As an ultimate long-term goal, mechanical sensing could open the possibility of direct detection of particle dark matter \emph{purely through its gravitational interaction with visible matter} \cite{PhysRevD.99.023005,PhysRevD.98.083019,Carney:2019pza}. This coupling is the only one guaranteed to exist, so an experiment with sufficient sensitivity would have the ability to find or completely rule out any dark matter candidate in the mass range for which it is sensitive. This proposal involves the direct monitoring of impulses delivered to sizeable (gram-scale) mechanical sensors, and exploits the coherent nature of the gravitational interaction. Achieving this goal would require realizing noise levels well below the SQL impulse sensing limit, as well as the ability to build and read out a large array of sensors. However, the concept employed is precisely the same as that described in the previous section, namely observation of an impulse to the center of mass of an object. The basic idea can thus be tested in prototype experiments, for example \cite{Monteiro2020compositeDM}.

\renewcommand\arraystretch{1.5}
 \begin{table}[t]
 \tiny
\begin{flushleft}
	\begin{tabular}{|p{3.5cm}|p{1.3cm}|p{1.7cm}|p{1.2cm}|p{2.2cm}|p{5.6cm}|}
		\hline
		Physical device & Mass & Frequency & Temp.  & Quantum limit & Sensitivity, e.g. acceleration, strain, force...  \\ \hline
	\end{tabular}
	
	\vspace{5pt} 
			Resonant acoustic wave: 
	\vspace{1pt} \\
	\begin{tabular}{|p{3.5cm}|p{1.3cm}|p{1.7cm}|p{1.2cm}|p{2.2cm}|p{5.6cm}|}
		\hline
		
		BAW/Weber bar \cite{branca2017search}   & 1000 kg  & 1 kHz  & 4 K  & & $h_s \sim 10^{-21}/\sqrt{\rm Hz}$                  \\ \cline{1-1} \hline	
		HBAR/phonon counting \cite{chu2017quantum}  & 50 $\mu$g & 10 GHz  &  10 mK & single phonon  &    \makecell[lt]{ $\sigma_E \sim 30\ \mu$eV   \\ $ h_s \sim 10^{-15}/\sqrt{\rm Hz} $ \\ $(h_s \sim 10^{-9}/\sqrt{\rm Hz} {\rm broadband below res}) $}  \\ \cline{1-1} \hline
		superfluid helium cavities \cite{shkarin2019quantum}  & 1 ng  & 300 MHz &  50 mK   & single phonon &     \makecell[lt]{  $\sigma_E \sim 1\ \mu$eV }       \\ \cline{1-1} \hline
		
	\end{tabular}
	
		\vspace{5pt} 
	Resonant and below-resonance detectors:
	\vspace{1pt} \\
		\begin{tabular}{|p{3.5cm}|p{1.3cm}|p{1.7cm}|p{1.2cm}|p{2.2cm}|p{5.6cm}|}
		\hline
		
		cantilever optomechanical accelerometer~\cite{guzman2014high}   & 25 mg   & 10 kHz & 300 K  &   &   \makecell[lt]{$\sqrt{S_{a}} \sim 3 \times 10^{-9}~\mathrm{g/\sqrt{Hz}}$ \\ ($\sqrt{S_{a}} \sim 10^{-7}~\mathrm{g/\sqrt{Hz}}$ broadband below res) }          \\ \cline{1-1} \hline
		SiN-suspended test mass accelerometer \cite{zhou2019testing,krause2012high}  & 10 mg   & 10 kHz &  300 K &   &  \makecell[lt]{ $\sqrt{S_{a}} \sim 10^{-7}~\mathrm{g/\sqrt{Hz}}$ \\  ($\sqrt{S_{a}} \sim 10^{-6}~\mathrm{g/\sqrt{Hz}}$ broadband below res)}        \\ \cline{1-1} \hline
		membrane optomechanics \cite{kampel2017improving, mason2019continuous, underwood2015measurement,tsaturyan2017ultracoherent,norte2016mechanical,reetz2019analysis,st2019swept}   & 10 ng   & 1.5 MHz &  100 mK & at SQL  &  \makecell[lt]{$\sqrt{S_{a}} \sim  10^{-7} \mathrm{g/\sqrt{Hz}}$  \\  $\sqrt{S_{f}} \sim 10^{-17}~\mathrm{N/\sqrt{Hz}}$   }    \\ \cline{1-1} \hline
		crystalline cantilever for force sensing \cite{mamin2001sub-attonewton} & 0.2 ng  & 1 kHz &  200 mK   &      &  \makecell[lt]{$\sqrt{S_{a}} \sim 3 \times 10^{-7} \mathrm{g/\sqrt{Hz}}$ \\ $\sqrt{S_{f}} \sim 10^{-18}~\mathrm{N / \sqrt{Hz}}$ }               \\ \cline{1-1} \hline

	\end{tabular}
	
	\vspace{5pt} 
	Pendula above resonance:
	\vspace{1pt} \\
	\begin{tabular}{|p{3.5cm}|p{1.3cm}|p{1.7cm}|p{1.2cm}|p{2.2cm}|p{5.6cm}|}
		\hline
		
		LIGO mirror \cite{martynov2016sensitivity}  &  10 kg   & 10 Hz -- 10 kHz  & 300 K & SN limited above 100 Hz  & \makecell[lt]{$\sqrt{S_{a}} \sim 4 \times 10^{-15}~\mathrm{g/\sqrt{Hz}}$ at 100 Hz \\ $\sqrt{S_{x}} \sim 10^{-19}~\mathrm{m/\sqrt{Hz}}$}   \\ \cline{1-1} \hline
		suspended mg mirror \cite{corbitt2007all-optical,matsumoto2019demonstration,catano2019high}  & 1 mg   & 1 -- 10 kHz  & 300 K & factor of 20 in displacement from (off-resonant)  SQL  &  \makecell[lt]{$\sqrt{S_{a}} \sim 7 \times 10^{-11} ~\mathrm{g/\sqrt{Hz}}$ at 600 Hz \\ $\sqrt{S_{x}} \sim 5 \times 10^{-17}~\mathrm{m/\sqrt{Hz}}$   }         \\ \cline{1-1} \hline
		crystalline cantilever \cite{cripe2019measurement} & 50 ng  & 10 -- 100 kHz   & 300 K  & at (off-resonant) SQL  & \makecell[lt]{$\sqrt{S_{a}} \sim 2 \times 10^{-7}~\mathrm{g/\sqrt{Hz}}$ at 20 kHz \\ $\sqrt{S_{x}} \sim 10^{-16}~\mathrm{m/\sqrt{Hz}}$ }           \\ \cline{1-1} \hline
		
	\end{tabular}

	\vspace{5pt} 
	Levitated and free-fall systems:
	\vspace{1pt} \\
		\begin{tabular}{|p{3.5cm}|p{1.3cm}|p{1.7cm}|p{1.2cm}|p{2.2cm}|p{5.6cm}|}
		\hline
		
		LISA pathfinder~\cite{anderson2018experimental}  &  15 kg   & 1 -- 30 mHz  & 300 K &  & $\sqrt{S_{a}} \sim 10^{-15}~\mathrm{g/\sqrt{Hz}}$   \\ \cline{1-1} \hline
		mm magnetically-levitated sphere \cite{timberlake2019magnetic}  &  4 mg  & 20 Hz  & 5~K  &   &   \makecell[lt]{ $\sqrt{S_{a}} \sim 2 \times 10^{-7}~\mathrm{g/\sqrt{Hz}}$ \\ $\sqrt{S_{f}} \sim 8 \times10^{-12}~\mathrm{N / \sqrt{Hz}}$     }  \\ \cline{1-1} \hline
		sub-mm magnetically-levitated sphere \cite{Lewandowski:2020cuq}  &  0.25 $\mu$g  & 1--20 Hz  & laser cool to $<9$~K  &   &  \makecell[lt]{$\sqrt{S_{a}} \sim 10^{-7}~\mathrm{g/\sqrt{Hz}}$  \\ $\sqrt{S_{f}} \sim 2 \times10^{-16}~\mathrm{N / \sqrt{Hz}}$  } \\ \cline{1-1} \hline	
		optically trapped microsphere \cite{monteiro2020optical}  &  1 ng  & 10 -- 100 Hz  &  laser cool to 50~$\mu$K  & factor of 100 in displacement from (off-resonant)  SQL  &  \makecell[lt]{$\sqrt{S_{a}} \sim 10^{-7} ~\mathrm{g/\sqrt{Hz}}$ \\ $\sqrt{S_{f}} \sim 10^{-18}~\mathrm{N / \sqrt{Hz}}$     }  \\ \cline{1-1} \hline	
		optically trapped nanosphere \cite{delic2019motional,tebbenjohanns2020optical} (rotational \cite{2020NatNa..15...89A}) &  3 fg  & 300 kHz  & laser cool to 12 $\mu$K   & ground state  & \makecell[lt]{ $\sqrt{S_{a}} \sim 7 \times 10^{-4}~\mathrm{g/\sqrt{Hz}}$ \\ $\sqrt{S_{f}} \sim  2\times10^{-20}~\mathrm{N / \sqrt{Hz}}$ \\ $ \sqrt{S_\tau} \sim 10^{-27}~\mathrm{N m/\sqrt{Hz}}$  }  \\ \cline{4-6}  \hline	
		trapped ion crystal \cite{biercuk2010ultrasensitive}  & $10^{-6}$ fg  & 1 MHz &     &     &  \makecell[lt]{$\sqrt{S_{a}} \sim 50~\mathrm{g/\sqrt{Hz}}$   \\ $\sqrt{S_{f}} \sim 4\times10^{-22}~\mathrm{N / \sqrt{Hz}}$    }           \\ \cline{1-1} \hline

	\end{tabular}
\end{flushleft}

	\caption{Examples of currently-available mechanical sensors. Sensitivities for continuous sensing are represented by the relevant noise power spectral densities (e.g.\ $S_a$ is the acceleration noise power), or threshold ($\sigma_E$ is the single-phonon detection threshold).  Here we summarize solid-state mechanical detectors, although atom interferometers can be characterized by similar metrics.}
		\label{device_table}
\end{table}
 
\section{Available mechanical sensors and future challenges}
\label{section-detectors}

Mechanical devices have been demonstrated with masses from single ions to kilograms, and on frequency scales from millihertz to terahertz. Precision sensing has long used massive detectors in the context of gravitational wave searches employing interferometric or resonant detectors, e.g.~LIGO.  On a smaller scale, accelerometers and other mechanical devices are ubiquitous in modern technology, and increasingly specialized mechanical systems with extreme environmental isolation are important tools for storage and transduction of quantum information~\cite{aspelmeyer2014cavity}.

As discussed above, many of the scientific motivations favor larger volumes or masses to increase the rate of dark matter interactions in the detector. This motivates use of more massive systems, which also provide better sensitivity to accelerations (scaling as the square root of the mass). However, also important are the energy range of interest, the available probes of specific mechanical modes, ever-present noise sources, and scalability.  To understand the scope of different available platforms, we present in Table~\ref{device_table} different detector types and a sampling of sensitivities achieved to date in specific experiments. This list is meant to be exemplary, and not exhaustive.  It can also be considered a starting point, i.e.\ rapid progress in mechanical detectors is being made in many fields, and as exemplified in the workshop on which this white paper is based, there is increasing cross-development between sensors of widely differing scales that will lead to fruitful technical improvements.

A central issue is to map the advantages of different physical architectures to different searches. For cases where an impulse detector is desired, an essentially free mass can be created by using a low-frequency pendulum measured above its resonance frequency, i.e. at time-scales faster than an oscillation period.  An interesting alternative is to levitate particles and then release them after state preparation to perform measurements in free-fall.  Ultralight searches are likely to be first pursued by resonant detectors---ideally tunable resonant detectors. The center of mass motion of a cantilever, membrane \cite{Manley:2020mjq}, or even levitated sphere are appropriate in this situation.  For ultralight searches that result in changes in atomic strain due to effective signatures that appear as time-variations in fundamental constants or atomic length scales, and hence excitation of effective breathing modes, bulk acoustic modes are of interest~\cite{manley2019searching}.  Importantly, detection of such bulk acoustic waves may scale to large volumes using clever readout techniques, as exemplified by recent single-phonon detection of a bulk acoustic resonator~\cite{chu2017quantum}, and in the long-standing ability to read out motion of very large Weber bars~\cite{weber1966observation,cerdonio1997ultracryogenic}.  Athermal phonon detection may also benefit from this scaling if athermal phonons created in the bulk of a material could be coupled into the readout modes of interest, but could also be pursued in arrays of smaller sensors.  Different devices can also support detection of additional signatures or couplings, e.g.\ electric or magnetic charges or the material polarizability.

The quest to go beyond the sensitivities presented in Table~\ref{device_table} is ongoing, and we list here a few examples of how advances in both conventional and non-conventional technologies for precision sensors are poised to make interesting progress.  Superfluid helium is a pristine system that hosts mechanical modes; recent advances \cite{shkarin2019quantum} in observing the quantum motion of this liquid in a small cavity are promising, and this system could be easily scalable to larger volumes and number of samples by simply immersing more probes in a single vat of liquid helium.  SiN micromechanical membranes offer a unique possibility to use strain to move the resonant frequency of a mechanical detector by orders of magnitude while maintaining low dissipation~\cite{thompson2008strong}, allowing searches over a wide range of DM masses.  By expanding to larger membranes \cite{moura2018centimeter,Manley:2020mjq} it should be possible to achieve kHz-scale resonant detectors with much larger masses than traditional cantilevers.  While optical readout is typical of precision interferometry, electrical readout is poised to make important contributions, both in the context of phonon readout through superconducting qubits~\cite{chu2017quantum}, but also through advances in magnetic couplings~\cite{zoepfl2019single}.  Detection of the motion of levitated nanospheres is reaching quantum measurement limits~\cite{delic2019motional}.  Scaling the mass of levitated systems in the quantum regime to the ng scale and above may offer extremely low threshold mechanical sensors with substantial mass that are well-isolated from environmental noise \cite{childress2017cavity,monteiro2020optical,timberlake2019magnetic}. Readout of ultra low-energy phonons is currently achieved in small devices; if these techniques could be adapted to read out larger volumes---and if the challenging problem of coupling energy from such a volume into the modes of interest could be overcome---the potential gains are significant. Lastly, the growth of gravitational wave astronomy will undoubtedly bring advances in materials for mirrors, mirror coatings, and suspensions that will advance all precision measurements based upon suspended pendula.

Reducing both technical and quantum measurement-added noise sources will allow for progressively increasing sensitivity to dark matter. In general, devices operating at lower frequencies tend to be dominated by thermal or other technical noise sources, while higher-frequency devices are limited by shot noise or more generally by quantum measurement noise. For systems in a $10~{\rm mK}$ dilution refrigerator, for example, the cutoff is at $\omega \sim k T/\hbar \sim 1~{\rm MHz}$. The primary contaminant in a dark matter search is the heating rate of a sensor, $\Gamma \sim T_{\rm bath}/Q$, where $Q$ is the mechanical quality factor. Thus fabrication of lower dissipation (higher-$Q$) devices will be of critical importance. 

We can see directly in Table~\ref{device_table} that a range of experiments are now impinging on quantum noise limits, and so methods to operate devices well into the quantum-limited regime (i.e.\ true ``quantum sensors'') are of substantial interest. Measurement-added noise has been suppressed below the shot noise limit at LIGO~\cite{abadie2011gravitational}, and it has likewise been driven to the standard quantum limit~\cite{kampel2017improving,cripe2019measurement} and beyond~\cite{mason2019continuous} with membranes and cantilevers.   Quantum sensing techniques can further reduce these noise levels using squeezed readout light \cite{aasi2013enhanced,tse2019quantum} and/or a variety of backaction-evasion techniques \cite{braginsky1980quantum,pereira1994backaction,clerk2008back,Ghosh:2019rsc}. In the context of free-mass targets, nanogram levitated spheres have been cooled to their quantum ground state~\cite{delic2019motional}. Ultimately, to detect momentum transfers far below the SQL, it may be necessary to prepare the mechanics in a more extreme non-classical state, such as a coherent spatial superposition, and then perform interferometric measurement \cite{geraci2015sensing,wan2016free,pino2018chip}. The sensitivity of such superpositions to small impulses is in principle unbounded, scaling with the spatial extent and temporal duration of the quantum coherence that is achieved.  In addition to sub-SQL sensitivities to classical forces, such an approach can offer the unique possibility of detecting sources of anomalous test-mass diffusion (e.g., DM-induced Brownian motion), which can cause decoherence in a matter interferometer \cite{riedel2013direct,riedel2017decoherence} even when the mean momentum transfer is negligible \cite{riedel2015decoherence}.

Construction and operation of an \textit{array of mechanical sensors} poses an interesting technical challenge with applications to many of the dark matter searches described above. Performing differential measurements on multiple sensors would allow for rejection of many backgrounds. In particular, use of sensors with different materials will enable discrimination against signals which act in a material-independent fashion, for example gravitational noise. Relative accelerations between objects with different numbers of neutrons could identify ultralight fields coupling to $B-L$. Coherent integration of multiple sensors would be highly valuable, enabling scaling in sensitivity that is linear with the number $N$ of sensors as opposed to the incoherent $\sqrt{N}$ enhancement. Understanding the detailed nature of sensor-sensor interactions in a tightly packed array will be important. These interactions could be exploited to enhance measurement sensitivity, in particular through entanglement of multiple sensors \cite{giovannetti2004quantum}.

In the near term, a number of demonstrator experiments could pave the way for future, scalable dark matter detection. Given the current constraints on ultralight dark matter, current or near future devices could already perform non-trivial searches in this parameter space. Operating a small array of sensors as a coherent detector of ultralight dark matter would demonstrate the basic techniques needed as well as help to identify challenges in scaling to larger numbers. Moving toward detection of short impulses, demonstration of ultra-low threshold phonon readout in a meaningful volume would be of substantial value. Demonstrating that optomechanical impulse sensing allows for backaction noise evasion would likewise be extremely valuable, and allow for a more detailed understanding of the potential limitations of such a technique, in particular due to optical losses.

\section{Conclusions}

Dark matter constitutes one of the most fundamental mysteries in modern science: what is the nature of this strange mass, taking up a quarter of the universe's energy budget? As the search for dark matter enters maturity, new theoretical and experimental directions are needed. Mechanical sensing technologies, especially with quantum-sensing techniques that can enable measurement past traditional quantum limits, offer an exciting route to new experimental searches.

Deploying currently available technology could have immediate impact, while longer-term prospects will require some technical advances. On the experimental side, a number of basic technological challenges to be overcome and demonstrations of the core search techniques will be of critical importance. Data processing techniques and the application of lessons learned from previous experiments about the nature of potential background signals will require development tailored to these experimental approaches. Looking toward the longer term, interdisciplinary collaborative efforts and the construction and use of multiple sensors as a coherent detector offer a fascinating set of problems.

Overall, the wide variety of platforms and scales available with these techniques has the potential to make significant impact across a wide swath of the dark matter landscape. Future developments should only continue to improve sensitivities and detection reach. Further collaboration between the mechanical quantum sensing and particle physics communities will undoubtedly lead to even more possibilities than those outlined here.

\section*{Acknowledgements}

We thank Charles W. Clark, Yiwen Chu, Tom Lebrun, and Jon Pratt for comments, and Yoni Kahn and Masha Baryakhter for suggesting the relevance of the weak gravity conjecture in Fig.~\ref{figure-ultralightplot}.  Yogesh S. S. Patil, Lucy Yu, and Sean Frazier produced the images in Fig.~\ref{figure-phonons}. This white paper originated with a workshop held at the Joint Quantum Institute at the University of Maryland, October 28-29, 2019. This workshop was funded in part by the Gordon and Betty Moore Foundation, through Grant GBMF6210. We also gratefully acknowledge support from  the JQI (an NSF Physics Frontier Center, award number 1430094), and from JILA (an NSF PFC, award number 1734006) to run the workshop. We thank the Aspen Center for Physics for hospitality during the workshop ``Quantum Information and Systems for Fundamental Physics'', where part of the writing was completed.

\bibliographystyle{utphys-dan}
\bibliography{references}

\end{document}